\documentclass[fleqn,usenatbib]{mnras}

\usepackage{newtxtext,newtxmath}

\usepackage{enumitem}
\setlist[enumerate]{leftmargin=*,labelindent=0pt}

\usepackage[T1]{fontenc}

\DeclareRobustCommand{\VAN}[3]{#2}
\let\VANthebibliography\thebibliography
\def\thebibliography{\DeclareRobustCommand{\VAN}[3]{##3}\VANthebibliography}

\usepackage{graphicx}	
\usepackage{amsmath}	
\usepackage{mathtools}

\usepackage{enumitem}
\usepackage{xcolor}
\usepackage{subcaption}

\title[Multi-messenger Analysis Of SMBHBs]{Multi-messenger Analysis of Supermassive Black Hole Binaries: The Joint-likelihood Approach}

\author[M.~Charisi et al.]{Maria Charisi,$^{1,2}$\thanks{E-mail: maria.charisi@nanograv.org} Stephen~R.~Taylor,$^{3}$
Jessie Runnoe,$^{3,4}$ 
Caitlin Witt,$^{5}$ 
Polina Petrov$^{3}$
\\
$^{1}$Department of Physics and Astronomy, Washington State University, Pullman, WA 99163, USA\\
$^{2}$Institute of Astrophysics, FORTH, GR-71110, Heraklion, Greece\\
$^{3}$Department of Physics \& Astronomy, Vanderbilt University, 2301 Vanderbilt Place, Nashville, TN 37235, USA\\
$^{4}$Fisk University, Department of Life and Physical Sciences, 1000 17th Avenue N, Nashville, TN 37208, USA \\
$^{5}$Wake Forest University, Department of Physics, Winston-Salem, NC 27103
}

\pubyear{2021}

\begin{document}
\label{firstpage}
\pagerange{\pageref{firstpage}--\pageref{lastpage}}
\maketitle

\begin{abstract}
Supermassive black hole binaries (SMBHBs) formed in galaxy mergers are promising multi-messenger sources. They can be identified as quasars with periodic variability in electromagnetic (EM) time-domain surveys. The most massive of those systems can be detected by Pulsar Timing Arrays (PTAs) in the nanohertz frequency gravitional-wave (GW)  band. We present a method to simultaneously analyze EM lightcurves and PTA observations as a multi-messenger data stream. For this, we employ a joint likelihood analysis, in which the likelihood of the EM data and the PTA likelihood are multiplied. We test this approach by simulating 208 binary signals that can be detected both by the Rubin Observatory in the nominal ten-year survey and by a PTA dataset with a $\sim$30-year baseline, which resembles our expectations for a dataset of the International Pulsar Timing Array (IPTA) collaboration in $\sim$2035. We compare our multi-messenger analysis with analyses that take into account the EM and PTA data separately. We find that the joint likelihood approach results in improved parameter estimation with smaller percent errors compared to the distinct analyses that consider only EM or PTA data separately. Among the SMBHB parameters, the binary total mass and the orbital inclination show the greatest improvement. We also compare our multi-messenger pipeline with an analysis, in which the EM constraints are used as priors to the PTA analysis. We demonstrate that the joint likelihood approach delivers tighter constraints on all binary parameters, with systematically higher values of Kullback–Leibler divergence, which measures the deviation of the posterior distribution from the prior.

\end{abstract}




\section{Introduction}
\label{sec:intro}

Supermassive black hole binaries (SMBHBs) should be fairly common in the universe, forming naturally in galaxy mergers \citep{Begelman1980}. They are among the most promising multi-messenger sources \citep{2019BAAS...51c.490K, Charisi2022}, since they are expected to emit both bright electromagnetic (EM) signatures \citep{2021arXiv210903262B,SMBHB_Review_Obs} and low-frequency gravitational waves (GWs). The most massive SMBHBs, with a mass of $10^8-10^{10}M_{\odot}$, emit nanohertz GW frequencies and can be detected by Pulsar Timing Arrays (PTAs; e.g., see \citealt{2021arXiv210513270T, Burke-Spolaor2019} for reviews), while the least massive binaries, with mass of $10^4-10^{7}M_{\odot}$, emit millihertz frequencies, which will be detectable in the future by the Laser Interferometer Space Antenna (LISA; see \citealt{LISA_review} for a review).

On the EM side, SMBHBs are expected to reach small separations surrounded by copious amounts of gas. The gas settles into a circumbinary disk and accretes onto the binary \citep{Barnes:2002}. Even though uncertainties regarding the effect of the gas on the binary evolution persist, several hydrodynamical simulations have converged on the following conclusion: SMBHBs can produce quasar-like luminosity, which is modulated periodically (see reviews by \citealt{2023ARA&A..61..517L,SMBHB_Review_Obs}). There are three main mechanisms that can produce periodic variability: (1) periodic mass accretion, (2) relativistic Doppler boost, and (3) self-lensing. 

More specifically, as the binary interacts with its gaseous surroundings, torques from the binary orbit expel the gas from the central region, creating a cavity of low-density material \citep{1994ApJ...421..651A}. This cavity is not completely devoid of gas because the binary pulls inward gaseous streams from the edge of the circumbinary disk \citep{1996ApJ...467L..77A}. The accretion rate is periodic, which likely translates into periodic variability \citep{farris15a,Tang+2018,westernacher-schneider22}. The periodicity is observed at the orbital period of the binary for unequal-mass binaries ($q<0.2$), whereas for roughly equal-mass binaries, the periodicity occurs at a few times (3-8) the orbital period \citep{dorazio13}. 

In addition, some of the gas that enters the cavity becomes bound to the individual SMBHs and forms persistent mini-disks \citep{Ryan2017}. The mini-disks move with velocities of a few percent the speed of light and relativistic effects become important. For unequal mass binaries with orbits not too far from edge-on, the variability is dominated by the rapidly moving secondary, which also has the most luminous mini-disk. The binary will appear brighter when the secondary is moving towards the observer and dimmer when it is receding, even if the rest-frame luminosity is constant \citep{Dorazio2015Nature}. When the orbit is roughly aligned with our line of sight, the luminosity of one mini-disk will be lensed from the other SMBH producing bright self-lensing flares, which repeat every orbit \citep{2018MNRAS.474.2975D, 2022PhRvL.128s1101D}.

Therefore, quasar periodicity is considered as one of the most promising EM signatures of SMBHBs \citep{2009ApJ...700.1952H}. In the last ten years, time-domain surveys have provided large samples of quasar lightcurves  and allowed for systematic searches, which have returned $\sim200$ promising candidates (see \citealt{SMBHB_Review_Obs} for a review). However, quasar variability is stochastic,
and can introduce false positives, especially when we cannot observe many cycles of periodicity within the available baselines \citep{Vaughan2016, Witt2022, Robnik2024}. As a result, these binary candidates have proven extremely challenging to confirm. Long-term monitoring and/or detecting additional signatures can boost our confidence in certain candidates. However, none of the proposed signatures is unique and in the absence of high-quality data, they can be confused with the variability of single-SMBH quasars \citep{2018MNRAS.476.4617C,SMBHB_Review_Obs}, making the detection of GWs almost a requirement for the confirmation of a SMBHB.

On the GW side, PTAs have recently opened the nanohertz window of the GW spectrum. All regional PTA collaborations, such as the North American Nanohertz Observatory for Gravitational Waves (NANOGrav; \citealt{McLaughlin2013,Ransom2019}), the European Pulsar Timing Array (EPTA; \citealt{2013CQGra..30v4009K}), along with the Indian Pulsar Timing Array (InPTA; \citealt{2022PASA...39...53T}), the Parkes Pulsar Timing Array (PPTA; \citealt{2008AIPC..983..584M,Hobbs2013}), the Chinese Pulsar Timing Array (CPTA; \citealt{2025A&A...695A.173X}) and the Meerkat Pulsar Timing Array (MPTA; \citealt{2023MNRAS.519.3976M}), have reported evidence for a stochastic low-frequency GW background \citep{NANOGrav_GWB,EPTA_GWB,PPTA_GWB,CPTA_GWB, MPTA_GWB}. This signal has consistent properties (amplitude, spectral slope) among different PTA datasets, albeit with different levels of significance \citep{IPTA_GWB_Comparison}. The source of this GW background is likely a population of unresolved SMBHBs \citep{NANOGrav_astro, EPTA_astro}, even though contributions from cosmological signals (e.g., inflation, phase transitions, cosmic strings, etc) are also possible \citep{NANOGrav_NewPhys}. 

In the next few years, individually resolved binaries should be detected on top of the GW background \citep{Rosado2015, Mingarelli2017, Kelley2018, Becsy2022}. This can enable multi-messenger discoveries, which offer great advantages. For example, targeting EM candidates can increase the detection sensitivity of PTAs \citep{Liu_Vigeland_MMA} or improve the GW upper limits \citep{2020ApJ...900..102A,2024ApJ...963..144A}. We also demonstrated that targeted searches can be more computationally efficient, since one can neglect the pulsar terms in the GW analysis without significantly affecting the parameter estimation \citep{Charisi2024}. 

In this paper, we present a simple proof-of-concept multi-messenger analysis. We jointly analyze EM time-domain data (e.g., from candidates identified as periodic quasars) and PTA data using a joint likelihood function. We emphasize that all previous multi-messenger analyses either fixed some of the GW parameters (e.g., the GW frequency) at the EM observed values or used the EM constraints as priors to the GW analysis. In our case, EM and GW data are analyzed as a single combined dataset.  With simulations, we explore how the parameter estimation is improved with the joint likelihood approach. We explore the scenario in which the periodicity is due to relativistic Doppler boost. In our previous study \citep{Charisi2022}, we showed that most of the binary parameters (e.g., the orbital period, the binary total mass, the mass ratio, orbital inclination) enter both the EM and GW likelihood function and can be linked in
the common likelihood.

The paper is structured as follows: in \S~\ref{sec:method} we summarize the binary simulations and the statistical analysis, and in \S~\ref{sec:results}, we present our results. We discuss caveats and future improvements in \S~\ref{sec:discussion}, while in \S~\ref{sec:summary}, we summarize our findings.

\section{Method}
\label{sec:method}

We simulate the following scenario. A SMBHB emits as a bright AGN with periodic brightness variations and is detected by the Rubin Observatory in the 10-year Legacy Survey of Space and Time (LSST). The same binary emits nanohertz GWs that can be detected by PTAs, which at the end of LSST will have a baseline of $\sim$30 years. In this scenario, the time-domain and PTA observations overlap during the last 10 years, although improvements are possible if the binary has archival time-domain observations before LSST. 

As we demonstrated in \cite{Charisi2022}, multi-messenger observations like the above scenario are possible for a variety of binaries. In fact, the binary parameter space that can be covered by both messengers expands as the PTA sensitivity increases. Since we explore the concept of jointly analyzing time-domain data and PTA data for the first time, in this proof-of-concept study, we make several simplifying assumptions. In a future study, we will address these limitations and conduct more realistic simulations (see \S~\ref{sec:discussion}).

\subsection{PTA Configuration}
We simulate a PTA dataset that resembles our expectations for an IPTA dataset in $\sim$2035. The dataset has a baseline of $\sim$30 years, and a total of 200 pulsars, each monitored for at least 3 years. For this, we follow the process described in \citet{Veronesi_2025}, expanding on previous simulations of PTA datasets by \cite{Pol_2021} and \cite{Petrov_2024}. In particular, our starting point is the upcoming 3$^{\rm rd}$ Data Release of the IPTA collaboration (IPTA DR3). Therefore, we start from a dataset, which consists of 116 pulsars, of which 68 are from the NANOGrav 15-year dataset \citep{2023ApJ...951L...9A}, 3 from the most recent data release of the EPTA, which also combines data from the InPTA \citep{2023A&A...678A..48E}, 14 from the 3$^{\rm rd}$ data release of PPTA \citep{2023PASA...40...49Z} and 31 from the 1$^{\rm st}$ data release of the MPTA \citep{2023MNRAS.519.3976M}. For pulsars monitored by multiple PTAs, we base our simulations on the NANOGrav timing data. 

We extend the observations of pulsars to achieve a total baseline of 30 years.
For the 45 pulsars that were included in the 12.5-year NANOGrav dataset, we add observations until we reach the 30-year baseline, following NANOGrav's observing strategy within the last year of observations (i.e. following the same observing pattern, with the same cadence, and uncertainties in the time of arrival, etc). For the remaining 71 pulsars that were not included in the 12.5-year NANOGrav dataset, we obtain their timing model parameters randomly from one of the following four pulsars: J0931-1902, J1453+1902, J1832-0836, and J1911+1347 (i.e. pulsars that were added in the NANOGrav 12.5-year dataset, but were not included in the NANOGrav 11-year dataset and are not in binary systems). We extend their timing data to reach a baseline of 30 years, adding bi-weekly observations and uncertainties in the time of arrival (TOA) equal to the white noise reported for each pulsar. 

Finally, we add 84 new pulsars, adding 7 pulsars per year, with each having at least 3 years of data when first included in the array. We randomly draw the sky coordinates of the new pulsars from the distribution of equatorial coordinates of the 116 monitored pulsars, which we approximate with a kernel density estimation (KDE) approach. We simulate the timing data, as in the 71  pulsars above (see \citealt{Veronesi_2025} for details).

\subsection{Binary TOA Simulations}
\label{subsec:TOA_Sims}
Next we inject a GW signal from a SMBHB with a circular orbit into each of these simulated PTA datasets, randomly drawing the binary parameters from uniform distributions in the following ranges:
\begin{itemize}
    \item \textbf{Sky Coordinates}, $\theta: [0, \pi]$, and $\phi: [0, 2\pi]$ \vspace{0.1cm}
    \item \textbf{Luminosity Distance}, $\log_{10} (D/\mathrm{Mpc}): [1,4]$ \vspace{0.1cm}
    \item \textbf{Total Binary Mass}, $\log_{10} (M_{\rm tot}/M_\odot):[9,10]$ \vspace{0.1cm}
    \item \textbf{Binary Mass Ratio}, $\log_{10}q: [-1,0]$ \vspace{0.1cm}
    \item \textbf{GW frequency}, $\log_{10} (f/\mathrm{Hz}):[-7.9,-7.2]$ \vspace{0.1cm}
    \item \textbf{Orbital Inclination Angle}, $\cos\iota:[-1,1]$ \vspace{0.1cm}
    \item \textbf{Initial Earth-term Phase}, $\Phi_0: [0,2\pi]$ \vspace{0.1cm}
    \item \textbf{GW Polarization Angle}, $\psi: [0,\pi]$
\end{itemize}

The simulated parameters are drawn from broad distributions, given our expectation for binaries that could be detectable with PTAs in the next 10 years \citep{Becsy2022, 2025arXiv250216016G, Veronesi_2025}. However, for the GW frequency, we also take into account that these binaries should also be detectable by LSST. Therefore, we set the maximum orbital period to 5 years (i.e. at least two cycles of periodicity are observed within the 10-year baseline), and thus the minimum GW frequency to $\log_{10}(f_{min}/Hz)=\log_{10}(2/P_{max})\sim-7.9$. Even though LSST is expected to detect binaries with shorter periods (albeit more rare), PTAs are unlikely to detect high frequency sources, and thus we set the minimum period to 1 year, which corresponds to a GW frequency of $\log_{10}(f_{max}/Hz)=\log_{10}(2/P_{min})\sim-7.2$.

The simulated timing deviations include the pulsar terms with pulsar distances from the respective datasets. 
The frequencies of the binaries do not evolve over the timing baseline of the data ($\sim$30 years), but we allow for frequency evolution between the Earth and the pulsar terms, which reflect the binary evolution thousands of years prior (since the pulsars are at kilo-parsec distances). Finally, we calculate the signal-to-noise ratio ($SNR$) for each injected binary, as in \cite{Petrov_2024} (e.g., see eq. 13). We keep in the sample binaries with $SNR$ between 5 and 15 to explore a range of possibilities in terms of signal strength. Out of 500 total simulations, 208 are consistent with the $SNR$ requirement, which comprise our final sample.

\subsection{Binary Lightcurve Simulations}
For each of the above combination of binary parameters, we simulate periodic lightcurves assuming that the variability of the binary is dominated by the relativistic Doppler boost of the secondary mini-disk, using eq. (33) in \cite{Charisi2022}.
These lightcurves are sinusoidal with a period equal to the orbital period of the binary. Their amplitude  depends on the inclination of the orbit, and the velocity of the secondary, which in turn depends on the total mass and mass ratio of the binary. The amplitude also depends on the
spectral index, which we fix at a fiducial value of $\alpha_\nu =-0.44$, obtained from the
composite spectrum of quasars \citep{VanderBeck}.
We note that even though the LSST data will be measured in magnitudes, here we simulate the lightcurves in fluxes. As we demonstrated in \cite{Charisi2022}, using fluxes is preferable for the multi-messenger data analysis, but the conversion between apparent magnitudes and fluxes is fairly straightforward.

We simulate lightcurves with a baseline of 10 years, and cadence of 5 days, similar to the Wide Deep Fast mode of LSST \citep{Ivezic_2019}, which will cover $\sim$95\% of the survey time. In these lightcurves, we do not include seasonal gaps, which are inevitable for ground-based observations. We consider single-band lightcurves, comparable to our expectations for LSST's $r$-band. We emphasize that LSST will rotate among six filters offering valuable multi-wavelength information, but combining the different bands is not trivial.
We include photometric errors drawing from a uniform distribution between 0.01 and 0.1. For simplicity, we ignore the underlying variability of quasars, which is typically described as a damped random walk \citep[DRW;][]{2009ApJ...698..895K}. Even though this can be a significant source of error, introducing false positives, it is unlikely to have a strong impact on the parameter estimation \citep{Witt2022}. As mentioned above, in this proof-of-concept study, we make several simplifying assumptions, which allow us to focus on the effects of the multi-messenger data combination, while avoiding limitations imposed by the data quality or the quasar noise. We plan to extend this study with more realistic simulations in the future (see \S~\ref{sec:discussion}).

\subsection{Statistical Analysis}
For each simulated binary dataset, which consists of a periodic lightcurve along with the timing deviations the SMBHB induces to the monitored pulsars in the simulated array, we perform four distinct Bayesian analyses:

\begin{enumerate}
    \item \emph{EM Only} analysis: We fit the lightcurves with a Doppler boost model.
    \item \emph{GW Only} analysis: We analyze the PTA data alone using an earth-term only model.
    \item \emph{GW + EM Prior} analysis: We repeat the PTA analysis, as before, but using priors informed from the lightcurve analysis.
    \item \emph{MMA} analysis:  We simultaneously analyze the EM and PTA data using a joint likelihood.
    
\end{enumerate}

As mentioned above, in the PTA analyses (both \emph{GW Only} and \emph{GW + EM Prior}), we neglect the pulsar terms, since, as we demonstrated in \citet{Charisi2024}, in targeted searches the inclusion of the pulsar terms only marginally improves the parameter estimation, while significantly increases the computational demands and complexity of the search. Below we briefly describe each individual analysis.

\subsubsection{EM Only analysis}
The EM lightcurve can be modeled with the following likelihood
\begin{equation}
     \ln p(d_\mathrm{EM}|\vec\theta_\mathrm{EM})  
     \propto -\frac{1}{2}\big[\ln \mathrm{det}(2\pi N) + (\vec{F_\nu} - \vec{s})^T N^{-1} (\vec{F_\nu} - \vec{s})\big]
\end{equation}
where $d_\mathrm{EM} \equiv \vec{F_\nu}$ are the flux measurements in the lightcurves, $N$ is a diagonal matrix with the photometric errors. The signal vector $\vec{s}$ is the Doppler boost model, dominated by the emission of the secondary
\begin{equation}
    s(t)=\left(3-\alpha_\nu\right)v_2/c\sin\iota\sin(\omega t+\Phi_0)
\end{equation}
where $\alpha_\nu$ is the spectral index of the AGN, which we keep fixed at the fiducial value, $\alpha_\nu =-0.44$ from \citet{VanderBeck},
$\omega=2\pi f_\mathrm{orb}$ is the orbital angular frequency (assumed to be approximately non-evolving over the observed baselines), $v_2$ is the velocity of the secondary SMBH, which depends on the total mass, $M=m_1+m_2$, the mass ratio, $q=m_2/m_1$, and the orbital period, $P_\mathrm{orb}$, of the binary (see eq. 5 in \citealt{Charisi2022}), $\iota$ is the orbital inclination angle measured between the binary angular momentum vector and a line-of-sight vector to the source, and $\Phi_0$ is a reference orbital phase measured at $t_0$.

\subsubsection{GW Only analysis}

The PTA likelihood and data analysis are described in great detail, in e.g., \cite{2014PhRvD..90j4012V,2021arXiv210513270T,NANOGrav_GWB}. Here we only present a summary of the salient points, but the reader is directed to the aforementioned references for a more in-depth technical description.

The PTA likelihood can be written as
\begin{equation}
    \ln p(d_\mathrm{PTA}|\vec\theta_\mathrm{PTA}) \propto -\frac{1}{2}\big[\ln \mathrm{det}(2\pi C) + (\vec{\delta t} - \vec{s})^T C^{-1} (\vec{\delta t} - \vec{s})\big],
\end{equation}
where $d_\mathrm{PTA}\equiv \vec{\delta t}$ is the concatenation of timing residuals from all pulsars in the array, which themselves are obtained through an initial fit of a deterministic timing model to each pulsar's TOAs. This fitting process is accounted for within the covariance matrix, which is defined as
\begin{equation} \label{eq:pta_covar}
    C = N + C_\mathrm{IRN} + C_\mathrm{GWB},
\end{equation}
where $N$ corresponds to diagonal in time (or block-diagonal in observing epochs) white noise contributions, $C_\mathrm{IRN}$ is low-frequency intrinsic red noise of each pulsar that is block diagonal in pulsars, and $C_\mathrm{GWB}$ is a dense matrix in which the GWB is modeled, with off-diagonal blocks modulated by the relevant Hellings \& Downs (or overlap reduction function) terms for those pulsar pairs. 
%

A continuous GW signal from an SMBHB is modeled entirely through $\vec{s}$, which is the concatenation of induced timing delays over all pulsars. In a single pulsar, $a$, this signal can be written as
\begin{equation}
    s_a(t) = F^+_a(\hat\Omega,\psi)\Delta s_+(t) + F^\times_a(\hat\Omega,\psi)\Delta s_\times(t),
\end{equation}
where $F^{\{+,\times\}}_a$ are GW antenna response functions for the $+$ and $\times$ GW polarizations, which depend on the sky location of the pulsar, the direction of GW propagation $\hat\Omega$, and (in this notation) the GW polarization angle $\psi$. The term $\Delta s_{\{+,\times\}}(t)\equiv s_{\{+,\times\}}(t_p) - s_{\{+,\times\}}(t)$ is the difference between the pulsar term and the Earth term, with $t_p = t - L_a(1+\hat\Omega\cdot\hat{p}_a)$ where $L_a$ is the distance to the pulsar and $\hat{p}_a$ is a unit vector pointing to the pulsar from Earth (or the Solar System barycenter). 
The respective time-dependent polarization terms to zeroth post-Newtonian order for a circular binary are
\begin{align}
    s_+(t) &= \frac{\mathcal{M}^{5/3}}{d_L\omega^{1/3}}(1+\cos^2\iota)\sin(2\omega t + 2\Phi_0), \nonumber\\
    s_\times(t) &= 2\frac{\mathcal{M}^{5/3}}{d_L\omega^{1/3}}\cos\iota\cos(2\omega t + 2\Phi_0),
\end{align}
where $\mathcal{M}$ is the (redshifted) binary chirp mass, with $\mathcal{M}~=~(1+z)M_\mathrm{tot}[q / (1+q)^2]^{3/5}$ 
and $d_L$ is the binary luminosity distance. It is typical for PTA searches for individual SMBHBs to set $z=0$, since PTAs can detect binaries only in the relatively nearby universe \citep{Veronesi_2025}. We follow this practice here as well.

Our PTA analyses are targeted, such that $\hat\Omega$ and $d_L$ are fixed to those of the EM candidates, leaving the binary mass, $M_{\rm tot}$, mass ratio, $q$, GW frequency, $f_{\rm gw}$, initial phase, $\Phi_0$, orbital inclination, $\iota$, and polarization angle, $\psi$, to be constrained by the PTA (or multi-messenger analyses below). 

\begin{figure*}
\includegraphics[width=\textwidth]{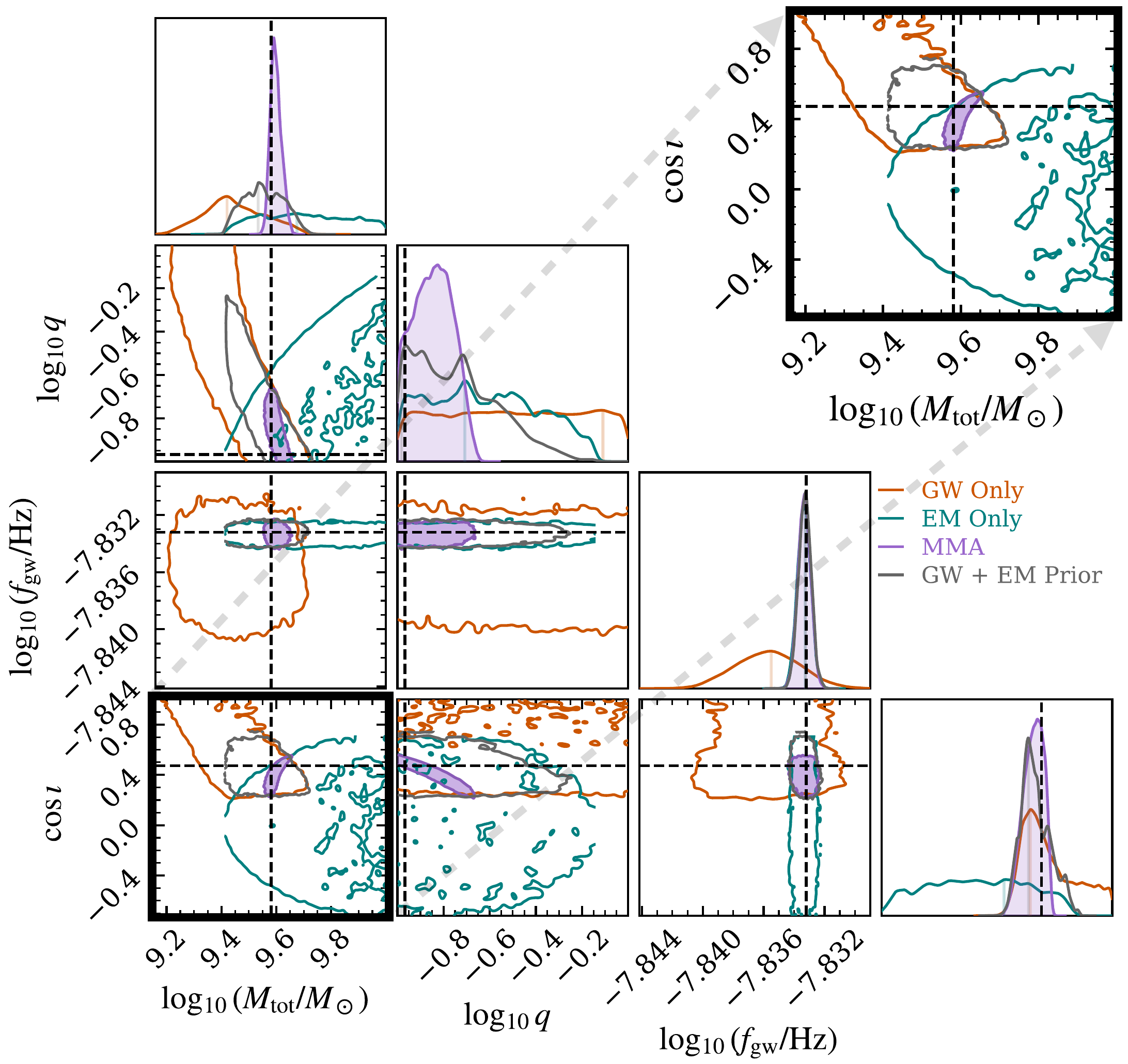}
\caption{Posterior distributions of the binary parameters, obtained with four different analyses. With dark orange we show the \emph{GW Only} analysis, dark green the \emph{EM Only} analysis, purple the \emph{MMA} and gray the \emph{GW + EM Prior} analyses, respectively. The dashed lines show the simulated parameters. We also show a zoom-in version of the 2D posterior distribution of the total mass and the orbital inclination, for which the \emph{MMA} analysis provided the tightest constraints. The simulated source for which this analysis was performed has $SNR\sim6$ in the 30-year IPTA dataset that we consider.}
\label{fig:cornerplot}
\end{figure*}

\begin{figure*}
\includegraphics[width=\textwidth]{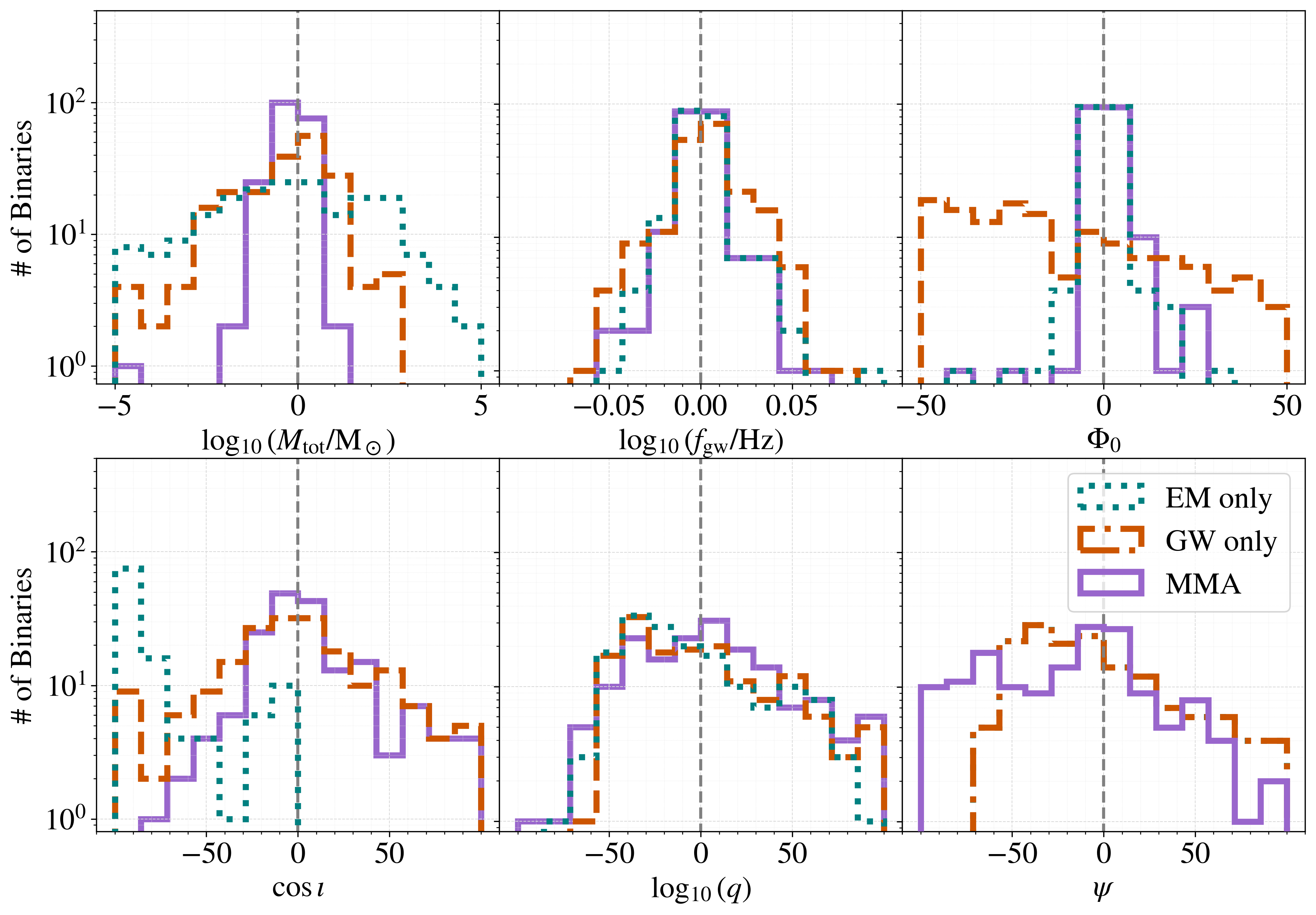}
\caption{Distributions of the percent error $\delta X[\%]$ for the six binary parameters we examine among the 208 binary simulations. With dotted dark green lines we show the \emph{EM Only} analysis, with dashed dark orange lines the \emph{GW Only} analysis, and with solid purple lines the \emph{MMA} analysis. Vertical dashed gray lines indicate 0\% error.}
\label{fig:percent_error}
\end{figure*}

\subsubsection{GW analysis with EM priors}
\label{subsub:GWEMPrior}
The simplest approach one could take to a multi-messenger analysis is to perform a Bayesian analysis on the EM data, recover the joint posterior of binary parameters, then convert the marginal posteriors of each parameter and use them as prior distributions for a subsequent analysis of the PTA data. This has been done before in \cite{Liu_Vigeland_MMA}. In our approach, we use Markov Chain Monte Carlo (MCMC) samples from an initial \emph{EM Only} analysis to create normalized KDE representations of the marginal posteriors of each binary parameter. This approach can be represented as
\begin{align}
    \ln p(\vec\theta_\mathrm{PTA})&|d_\mathrm{PTA}) \propto \nonumber\\
    & \ln p(d_\mathrm{PTA}|\vec\theta_\mathrm{PTA}) + \ln p(\vec\varphi_\mathrm{PTA}) + \sum_\alpha \ln p(\eta_\alpha|d_\mathrm{EM}),
\end{align}
where $\vec\varphi_\mathrm{PTA}$ are the parameters (e.g.,
noise, GWB, and even GW-specific binary parameters, like the polarization angle, $\psi$ that are not constrained by the \emph{EM Only} analysis), while $\eta_\alpha$ represents the parameters constrained through the \emph{EM Only} analysis.
The summation term implies that we are artificially factorizing the posterior constraints from the EM data into individual marginal terms. 

This two-step multi-messenger analysis has the benefit of being simple to implement, yet it has some clear drawbacks. It ignores covariances in the joint posterior distribution of binary parameters constrained by the EM data, and propagates this implicit assumption through to the PTA analysis. Albeit a step in the right direction, it can be considered only as an approximation of a true multi-messenger analysis. 
One could rectify this assumption while retaining a two-step analysis structure by reweighting the MCMC samples from the \emph{EM Only} analysis according to the PTA likelihood. However, this encounters an obvious issue -- parameters in $\vec\varphi_\mathrm{PTA}$ remain unconstrained. While it is possible to run separate EM and PTA analyses, and reweight samples appropriately after the fact, the most robust approach to a multi-messenger analysis is to construct a joint likelihood over the datasets, as described below.

\subsubsection{Multi-messenger analysis}
\label{subsub:MMA_Analysis}
Here we present a multi-messenger likelihood, which is simply the product of the likelihoods of the two messengers, with a parameter space that covers all the distinct parameters of the PTA, and EM models, $\vec\theta_\mathrm{PTA}$, and $\vec\theta_\mathrm{EM}$, respectively, such that 

\begin{equation}
\mathcal{L}_\mathrm{MMA}=\mathcal{L}_\mathrm{PTA}\mathcal{L}_\mathrm{EM},
\end{equation}
or more specifically,
\begin{equation}
\ln p(d_\mathrm{MMA}|\vec\theta_\mathrm{MMA}) = \ln p(d_\mathrm{PTA}|\vec\theta_\mathrm{PTA}) + \ln p(d_\mathrm{EM}|\vec\theta_\mathrm{EM}),
\end{equation}
where $\vec\theta_\mathrm{PTA} = (\vec\eta,\vec\varphi_\mathrm{PTA})$, $\vec\theta_\mathrm{EM} = (\vec\eta,\vec\varphi_\mathrm{EM})$, and $\vec\theta_\mathrm{MMA} = (\eta,\vec\varphi_\mathrm{PTA},\vec\varphi_\mathrm{EM})$, with $\vec\eta$  the vector containing all common parameters of the models that are constrained through the multi-messenger combination of the PTA and EM datastreams---$d_\mathrm{PTA}$ and $d_\mathrm{EM}$, respectively. In our case, these are the intrinsic parameters of the binary itself, $\{M_\mathrm{tot},q,f_\mathrm{orb},\cos\iota,\Phi_0\}$, whereas $\vec\varphi_\mathrm{PTA}$ and $\vec\varphi_\mathrm{EM}$ are all remaining noise and signal parameters that are present in only one of the factorized terms of the multi-messenger likelihood.

What remains is to apply appropriate priors (only once) to each parameter in $\vec\varphi_\mathrm{MMA}$. This is important to avoid double-counting prior terms. Upon factorizing the prior, the result is 
\begin{equation}
    p(\vec\theta_\mathrm{MMA}) \equiv p(\eta,\vec\varphi_\mathrm{PTA},\vec\varphi_\mathrm{EM}) = p(\vec\eta)p(\vec\varphi_\mathrm{PTA})p(\vec\varphi_\mathrm{EM}).
\end{equation}
Bayes' theorem then allows us to constrain the (unnormalized) joint posterior distribution of all parameters as
\begin{align}
    \ln p(\vec\theta_\mathrm{MMA}|d_\mathrm{PTA},d_\mathrm{EM}) \propto& \ln p(d_\mathrm{PTA}|\vec\eta,\vec\varphi_\mathrm{PTA}) + \ln p(d_\mathrm{EM}|\vec\eta,\vec\varphi_\mathrm{EM}) \nonumber\\
    &+ \ln p(\vec\eta) + \ln p(\vec\varphi_\mathrm{PTA}) + \ln p(\vec\varphi_\mathrm{EM}).
\end{align}

\begin{figure*}
\includegraphics[width=\textwidth]{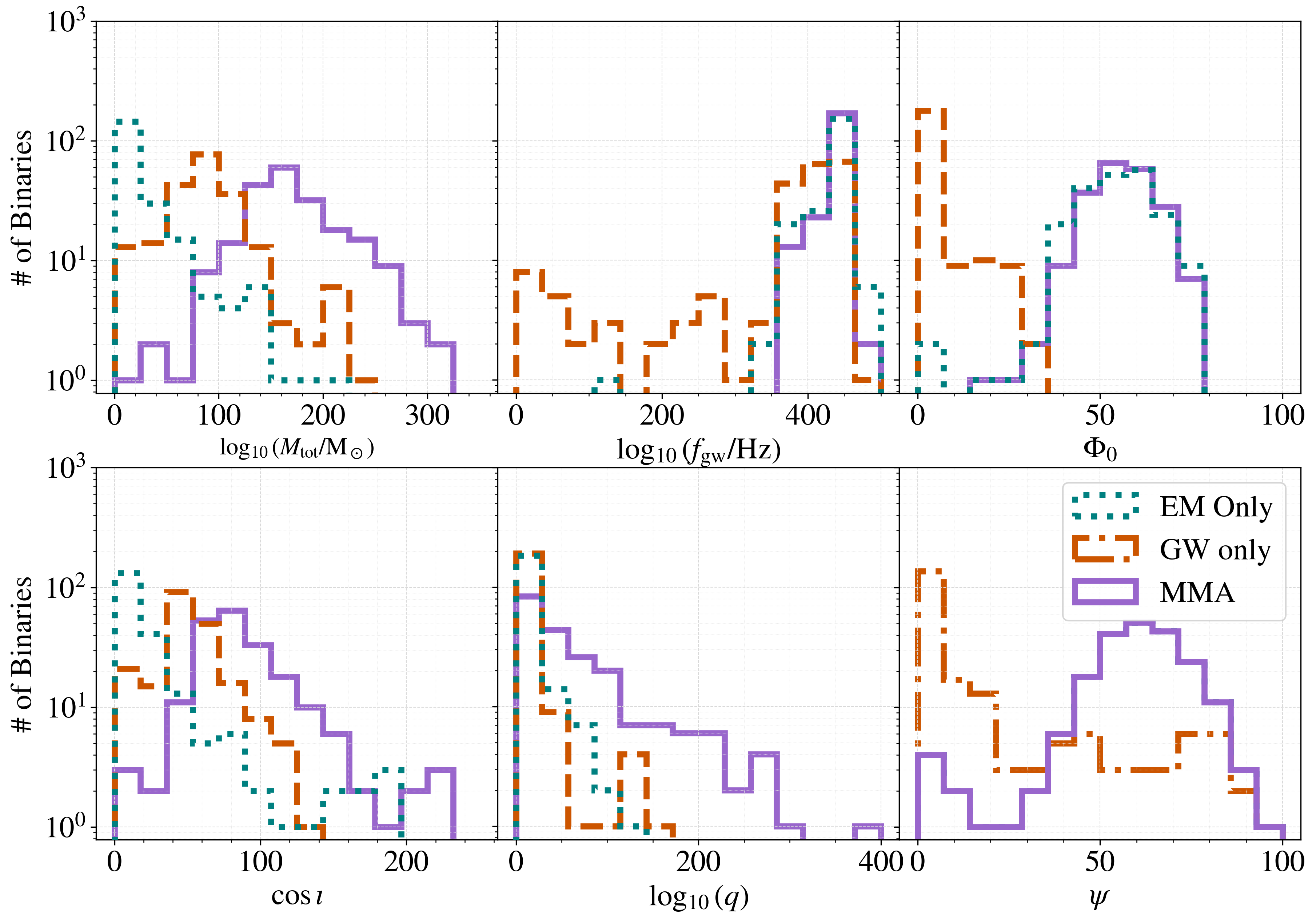}
\caption{Distributions of KL divergence for the six binary parameters we examine among the 208 binary simulations. The KL divergence quantifies the deviations observed in the posteriors compared to the priors, with high KL divergence values indicating that the data are informative, while low values indicate that the data are not very informative. We observe the highest KL divergence values for the GW frequency $f_{\rm gw}$, which is the best constrained parameter, and the lowest values for the polarization angle, $\psi$, which is hard to constrain. Color coding and line styles as in Figure \ref{fig:percent_error}.}
\label{fig:kl_divergence}
\end{figure*}

\subsubsection{Posterior Sampling}
For each of our simulations, we perform four separate analyses, as described above. We use the \texttt{enterprise} PTA software \citep{enterprise}, and \texttt{PTMCMCSampler} \citep{ptmcmc} to sample the different likelihoods we consider. We use uniform priors for the explored parameters, covering the same range as the injected parameters (see \S~\ref{subsec:TOA_Sims}),  except for the \emph{GW + EM Prior} analysis, where as detailed in \S~\ref{subsub:GWEMPrior}, we use EM-informed priors. 
In Figure~\ref{fig:cornerplot}, we show an example of the marginalized posterior probability distributions of four of the binary parameters (total mass,  $M_{\rm tot}$,  mass ratio, $q$, GW frequency, $f_{\rm gw}$, and orbital inclination, $\iota$) for a source whose GW signal has an $SNR\sim6$ in the 30-year IPTA dataset we consider. The dark orange contours represent the \emph{EM Only} analysis, the dark green the \emph{GW Only} analysis, while the gray and purple show the two multi-messenger analyses, i.e. the \emph{GW + EM Prior} and the \emph{MMA} analysis, respectively. For emphasis, we have shaded the contours of the \emph{MMA} analysis. We also show the injected values of the simulated binary with dashed lines. In a zoom-in inset, we highlight the 2D posterior distribution of the total mass and inclination, which are among the parameters that benefit the most from the \emph{MMA} analysis. Overall, we see that the \emph{MMA} analysis provides the tightest constraints on the parameters, followed by the \emph{GW + EM Prior} analysis. The other two analyses, in which either the lightcurve or the GW signal are considered individually (\emph{EM Only} and \emph{GW Only}, respectively), show larger uncertainty in the parameter estimation, while in some cases, the posteriors are not informative at all. For instance, the \emph{GW Only} analysis practically returns the prior distribution for the mass ratio, $q$ while the \emph{EM Only} analysis shows a very broad posterior for the inclination, $\iota$ (not very different from the prior).

\section{Results}
\label{sec:results}
We simulated EM and GW signals of 208 SMBHBs that could be detectable both by LSST and PTAs. In particular, we simulated idealized 10-year periodic lightcurves, in which the periodicity arises from relativistic Doppler boost of the secondary mini-disk. We also simulated the respective TOAs induced in a 30-year IPTA-like dataset.
Below we present the results for the sample of simulated binaries. We discuss how the parameter estimation improves with the joint likelihood analysis, compared to when the EM and GW data are analyzed separately. Then we carefully compare the results of the \emph{MMA} with the \emph{GW + EM Prior} analysis. 

\begin{figure*}
\includegraphics[width=\textwidth]{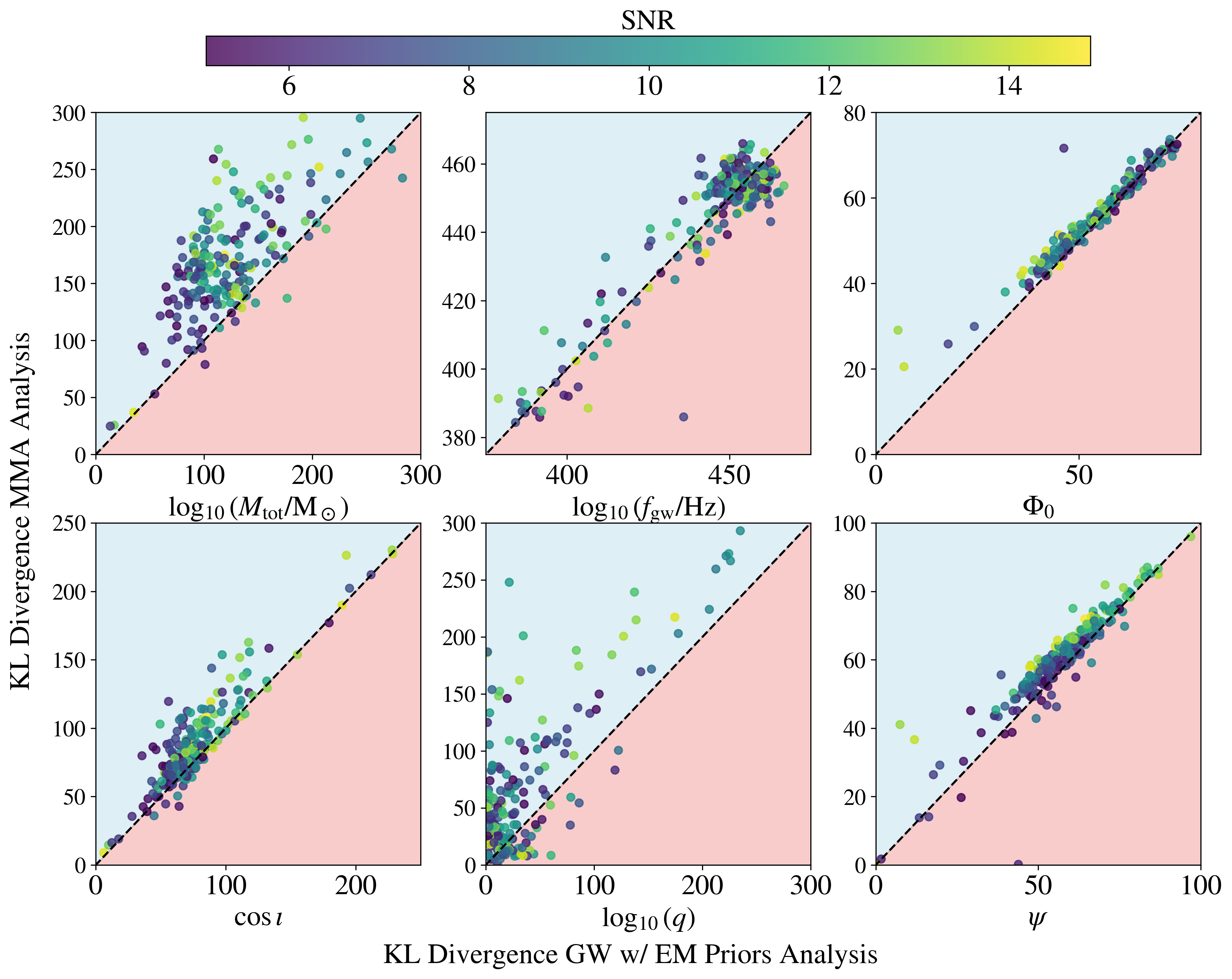}
\caption{Comparison of KL Divergence of the posteriors of the \emph{MMA} analysis versus KL divergence from the \emph{GW + EM Prior} analysis for each binary parameter we consider. Each point corresponds to a distinct simulation and is color-coded according to its $SNR$ in the 30-year IPTA-like array. Simulations that fall in the blue/red shaded regions of the plots, above/below the equality line obtain better constraints in the \emph{MMA}/\emph{GW + EM Prior} analysis.}
\label{fig:mma_comparison}
\end{figure*}

\subsection{EM Only and GW Only versus MMA analysis}

For the population of 208 binaries, we examine how well the injected values, $X_{\rm in}$, are recovered across all our simulated datasets, by calculating the percent errors of the posterior median, $X_{{\rm post},50}$, as:
\begin{equation}
    \delta_X [\%]=\frac{X_{\rm in}-X_{{\rm post},50}}{X_{\rm in}}\times100\%
\end{equation}
where $X$ is any of the six binary parameters. In Figure~\ref{fig:percent_error}, we show the distribution of the percent error for each binary parameter with color-coding similar to Figure~\ref{fig:cornerplot}, i.e. solid purple lines for the \emph{MMA} analysis, dashed dark orange lines for the \emph{GW Only} analysis and dotted dark green for the \emph{EM Only} analysis, respectively.
We see that all three analyses successfully constrain the GW frequency with $|\delta f_{\rm gw}|<1\%$ for the vast majority of simulations. We remind the reader that in the Doppler boost model we consider, the periodicity in the lightcurve corresponds to the orbital period of the binary, which can be directly linked to the GW frequency as $ f_{\rm gw}=2/P_{\rm orb}$. In addition, we see that the distributions for the total mass and inclination, $\delta M_{\rm tot}$ and $\delta \cos \iota$, are more peaked around $0\%$ in the \emph{MMA} analysis compared to the \emph{GW Only} analysis, and even more so compared to the \emph{EM Only} analysis, and thus the \emph{MMA} analysis provides better constraints for these two parameters.
The fact that the total mass and inclination are not well-constrained in the \emph{EM Only} analysis is not surprising, given that they both affect only the amplitude of the lightcurve, and hence they are degenerate. Among the other parameters, the initial phase, $\Phi_0$, is also fairly well constrained in the \emph{MMA} analysis, but very similar to the \emph{EM Only} analysis. This means that the \emph{MMA} constraints are likely driven by the lightcurve data, since in a sinusoisal lightcurve, it is fairly straightforward to fit the phase of the signal, while in the \emph{GW Only} case, the phase is not as easily constrained (e.g., see \citealt{Witt2022} for previous work on modeling EM lightcurves and \citealt{Charisi2024} for fitting PTA data). Finally, the mass ratio, $q$, and polarization angle, $\psi$, are only slightly more peaked around $0\%$ for the \emph{MMA} analysis compared to the other two, but generally show very broad distributions of percent errors indicating that these parameters are relatively hard to measure even with the \emph{MMA} analysis.

However, since the percent error, $\delta X [\%]$, reflects only the mean of the distribution, we also calculate the Kullback–Leibler (KL) divergence. The KL divergence measures how much the posterior distribution deviates from the prior, and thus allows us to capture 
how much information we gain from the data, taking into account the entire posterior distribution and not just a point estimate. A high value of KL divergence indicates significant deviation of the posterior from the prior, meaning that the data are informative, whereas if the KL divergence is close to zero, it means that the posterior and prior coincide, and thus the data cannot update our prior knowledge (see \citealt{Liu_Vigeland_MMA} and \citealt{Charisi2024} for more details). In Figure~\ref{fig:kl_divergence}, we present the distributions of KL divergence for all six binary parameters, using the same colors and line styles as in Figure~\ref{fig:percent_error}. We observe the highest values of KL divergence for the GW frequency $f_{\rm gw}$, which is the best constrained parameter in all analyses, with the \emph{MMA} and \emph{EM Only} analyses showing similar distributions and higher overall values compared to the \emph{GW Only} case. We find similar results for the initial phase, $\Phi_0$ since as we mentioned before, the constraints are likely driven by the EM data. For the remaining parameters, total mass, $M_{\rm tot}$, mass ratio, $q$ and inclination, $\cos \iota$, the \emph{MMA} analysis returns higher values of KL divergence compared to the other two analyses, indicating that these parameters are better constrained with the joint likelihood approach. Finally, for the polarization angle, $\psi$, the gain in information is relatively low (compared to the other parameters), but higher for the \emph{MMA} analysis compared to the \emph{GW Only}. We conclude that the \emph{MMA} analysis provides overall better constraints compared to analyses in which only one messenger is taken into account.

\subsection{Comparison of the two multi-messenger analyses}
Here we compare the two different possibilities of performing a multi-messenger analysis, the \emph{GW + EM Prior}, described in \S~\ref{subsub:GWEMPrior}, in which the EM lightcurve is analyzed first and the posteriors of this analysis are used as priors for the PTA analysis, and the \emph{MMA} analysis, described in \S~\ref{subsub:MMA_Analysis}, which employs a joint likelihood and simultaneously analyzes EM and PTA data. 

In Figure~\ref{fig:mma_comparison}, we compare the KL divergence measured from the posteriors of the \emph{MMA} analysis versus the respective values from the \emph{GW + EM Prior} analysis for all binary parameters and all binary simuations. We color code the binaries according to the $SNR$ of the GW signal in the 30-year IPTA dataset. Points that fall in the blue shaded region of the figure correspond to simulations where the \emph{MMA} analysis returns higher values of KL divergence compared to the \emph{GW + EM Prior} analysis, while points in the red shaded region show lower KL divergence values. Overall, we see that the KL divergence tends to be higher for most simulations and binary parameters in the \emph{MMA} case, which demonstrates that the parameter constraints obtained with this method are more informative compared to the \emph{GW + EM Prior} analysis. 

More specifically, the KL divergence for the GW frequency, $f_{\rm gw}$, shows similar and overall high values in both analyses. For the initial phase, $\Phi_0$, and the polarization angle, $\psi$, the points are clustered along the equality line, but with $\sim 80-85\%$ of the simulations showing slightly higher KL divergence in the \emph{MMA} analysis. The highest gain of information is seen in the total mass, $M_{\rm tot}$, with $\sim92\%$ of simulations showing higher numbers of KL divergence in the \emph{MMA} analysis. We see similar improvements for the inclination, $\cos \iota$, and the mass ratio, $q$, with $82\%$ and $73\%$ of simulations having higher KL divergence in the \emph{MMA} analysis, respectively.
Finally, we do not observe strong trends with the $SNR$, but stronger signals tend to have somewhat higher KL divergence values, i.e. their parameters are better constrained. We conclude that the joint likelihood method performs better than the \emph{GW + EM Prior} option in constraining the binary parameters.

\section{Discussion}
\label{sec:discussion}
We present a multi-messenger approach to analyze EM time-domain data (i.e. periodic lightcurves of quasars) together with PTA data of SMBHB candidates. With simulations, we demonstrate that the joint likelihood approach proposed in this study (\emph{MMA} analysis) can be advantageous, providing better parameter constraints compared to analyzing the EM or the PTA data individually (\emph{EM Only} and \emph{GW Only} analyses) or using the EM constraints as priors for the GW analysis (\emph{GW + EM Prior} analysis), which offers an alternative option for a multi-messenger pipeline \citep{Liu_Vigeland_MMA}. These results are very encouraging, but, in this first proof-of-concept study, we have made some simplifying assumptions. In a follow-up study, we plan to address these limitations to further examine the robustness of our conclusions.

First, for the EM signature of the binary, we assume that the periodicity arises from relativistic Doppler boost dominated by the emission of the secondary mini-disk. We chose this signature, because as discussed in \citet{Charisi2022}, this is the most promising scenario for combining EM with PTA data; the observed period corresponds to the orbital period of the binary, and thus can be linked to the GW frequency, $f_{\rm gw}$, while the amplitude of the periodic signal is determined by several of the binary parameters that also affect the GW signal (total mass, $M_{\rm tot}$, mass ratio, $q$, and orbital inclination, $\cos \iota$). 
Even though the relativistic Doppler boost is likely a prominent mechanism for periodic variability, it is not the only possibility (see \citealt{SMBHB_Review_Obs} for a review).
Periodic variability of SMBHBs may also arise from periodic modulations in the accretion rate. 
In this scenario, the observed periodicity may coincide with the orbital period of the binary (for unequal mass binaries) or may be a few (3-8) times longer (for higher mass ratios), corresponding to the orbital period of a hotspot that forms in the circumbinary disk. Therefore, the connection of the EM periodicity with the GW frequency, $f_{\rm gw}$, is more tentative compared to the Doppler boost case. In addition, it is not straightforward to connect the amplitude of the observed periodicity to any of the binary properties.
The above would likely have a negative impact on the \emph{MMA} analysis, since the EM data will not be equally constraining. On the other hand, for orientations close to edge-on, strong self-lensing flares can be present and repeat at the orbital period of the binary \citep{2018MNRAS.474.2975D}. Observing and modeling such flares in the EM lightcurve could provide strong constraints on the binary parameters, thus significantly boosting the \emph{MMA} parameter estimation.

Beyond the theoretical considerations of the binary signals, our future study will also include more realistic EM data. Here we simulate lightcurves with a nominal cadence of 5 days, resembling our expectations for $r$ band data in the Wide Fast Deep mode of LSST. However, our simulations are fairly idealized, since they do not include seasonal gaps, which are inevitable for ground-based surveys. For simplicity, we have also neglected the stochastic noise of quasars, which is typically described by a DRW model. Even though the red noise variability presents a great challenge in quasar periodicity searches, both introducing false positives and hindering the detection of real signals \citep{Vaughan2016, Witt2022,Robnik2024, 2025arXiv250514778L}, it does not significantly affect the parameter estimation \citep{Witt2022} and thus its inclusion is not expected to significantly alter our results. Moreover, LSST will observe in six distinct filters providing multi-band lightcurves. Including all the lightcurves in the \emph{MMA} analysis can be beneficial, especially for models like the relativistic Doppler boost, which are wavelength-dependent.

On the GW side, since we consider the parameter estimation in targeted multi-messenger searches, we perform an Earth-term analysis, which neglects the pulsar terms. In a previous study, we demonstrated that the Earth-term analysis provides similar parameter constraints for targeted GW searches, while avoiding the complexity and computational demands of the full signal analysis \citep{Charisi2024}. However, as we also showed in \citet{Charisi2024}, some of the binary parameters, e.g., the mass ratio, $q$, would likely improve with the inclusion of pulsar terms. We envision that future \emph{MMA} analyses will likely take a multi-tier approach, starting from the simpler joint likelihood, in which the GW analysis relies only on the Earth-term, like the one we examine here. Then for the most promising binaries, they will likely proceed with a full PTA likelihood, which includes the pulsar terms. In a future study, we will explore potential improvements of a multi-messenger pipeline, which takes into account the pulsar terms. Moreover, following standard practices in PTA searches for individual binaries, we have neglected the effects of redshift and eccentricity. In particular, we set $z=0$, since we expect that detectable sources are in the relatively nearby universe, i.e. $z<0.5$ \citep{Veronesi_2025}. We also consider only circular binaries; eccentric binaries show more complex periodic profiles and their GW signal is spread across multiple harmonics. It is unclear how the inclusion of eccentricity can affect the binary parameter estimation, but we will address this carefully in future work.

Finally, in this study we explored the \emph{MMA} analysis in the context of parameter estimation. It is also important to explore whether a joint likelihood analysis, like the one we consider here can boost the detectability of signals, as was found to be the case in the \emph{GW + EM Prior} analysis   \citep{Liu_Vigeland_MMA}. It is also likely that such an \emph{MMA} analysis could improve upper limits on binary candidate systems \citep{2020ApJ...900..102A,2024ApJ...963..144A, 2025arXiv250820007C} and could improve PTA searches, which target EM candidates \citep{2025arXiv250816534A}

\section{Summary}
\label{sec:summary}

In this paper, we present a novel method to simultaneously analyze EM time-domain and PTA data of SMBHB candidates. We propose a
joint likelihood approach, in which the EM likelihood and the PTA likelihood are multiplied. We test this method with 208 simulations of EM and PTA signals, examining LSST-like lightcurves, in which the periodicity comes from relativistic Doppler boost of the secondary, and their respective PTA signals induced in a 30-year IPTA-like dataset. 
We demonstrate that the proposed \emph{MMA} analysis, which employs the joint likelihood, outperforms the \emph{EM Only} and \emph{GW Only} analyses, where the lightcurves or the PTA data are analyzed separately, as well as the \emph{GW + EM Prior} analysis, in which the constraints from the EM analysis are used as priors for the PTA analysis. We compare the percent error of the posterior median and the KL divergence, which quantifies the deviations of the posterior distribution from the prior (i.e. the gain of information from the data).
Our results can be summarized as follows:
\begin{itemize}
    \item The GW frequency, $f_{\rm gw}$, is well constrained in all analyses, with percent error close to 0\% for the majority of simulations and high values of KL divergence. 
    \item The initial phase, $\Phi_0$, is well constrained in the \emph{MMA} analysis, driven by constraints from the EM data, while the \emph{GW Only} analysis struggles to constrain this parameter. The comparison with the \emph{GW + EM Prior} analysis provides similar results, with 80\% of simulations showing slightly higher values of KL divergence in the \emph{MMA} analysis.
    \item The constraints on total mass, $M_{\rm tot}$, and orbital inclination, $\cos \iota$, improve the most from the \emph{MMA} analysis, showing distributions of percent error more peaked close to 0\% compared to the \emph{EM Only} and \emph{GW Only} analyses. Comparing the \emph{MMA} with the \emph{GW + EM Prior} analysis, we find that 92\% and 82\% of simulations, respectively, have substantially higher KL divergence, which means that the \emph{MMA} analysis provides tighter constraints on these two parameters.
    \item The mass ratio, $q$, and polarization angle, $\psi$, are only slightly better constrained in the \emph{MMA} analysis, but overall show wide distributions of percent errors and relatively low KL divergence values. 
\end{itemize}

Future work will extend this study to include more EM signatures (e.g., periodicity due to accretion modulations or self-lensing), more realistic EM data (including gaps and red noise), and improvements in the PTA data (e.g., including pulsar terms). We will also explore the effect of the \emph{MMA} analysis on the detectability of binaries beyond the parameter estimation.

\section*{Acknowledgements}
We thank our colleagues in NANOGrav and the International Pulsar Timing Array for useful discussions about this work. Special thanks to Nihan Pol, William Lamb and Kayhan Gultekin. We appreciate the support of the NANOGrav NSF Physics Frontier Center awards \#2020265 and \#1430284.  
MC acknowledges support by the European Union (ERC, MMMonsters, 101117624). SRT acknowledges support from NSF AST-2307719 and NSF CAREER \#2146016. SRT is also grateful for support from a Vanderbilt University Chancellor's Faculty Fellowship. JCR acknowledges support from NSF AST-2205719. PP acknowledges support from
NASA FINESST grant number 80NSSC23K1442. This research used resources of the Center for Institutional Research Computing at Washington State University.
This work was conducted in part using the resources of the Advanced Computing Center for Research and Education (ACCRE) at Vanderbilt University, Nashville, TN.


\section*{Data Availability}
No new data were generated or analysed in support of this research.


\bibliographystyle{mnras}
\bibliography{mma} 






\bsp	
\label{lastpage}
\end{document}